\documentclass[12pt]{iopart}

\usepackage{iopams}
\usepackage{amsfonts}
\usepackage{amssymb}
\usepackage{calc}
\usepackage{graphicx}
\usepackage{bm}
\usepackage{color}
\usepackage{citesort}

\def\be{\begin{equation}}
\def\ee{\end{equation}}
\def\bea{\begin{eqnarray}}
\def\eea{\end{eqnarray}}
\def\bse{\begin{subequations}}
\def\ese{\end{subequations}}

\begin{document}

\title{Simulation of Jahn-Teller-Dicke Magnetic Structural Phase Transition with Trapped Ions}

\author{Peter A. Ivanov}
\address{Department of Physics, Sofia University, James
Bourchier 5 Boulevard, 1164 Sofia, Bulgaria}
\address{Institut f\"ur Quantenphysik, Johannes
Gutenberg-Universit\"at Mainz, 55099 Mainz, Germany}
\author{Diego Porras}
\address{Departamento de Fisica Te\'orica I,
Universidad Complutense,
28040 Madrid,
Spain}
\author{Svetoslav S. Ivanov}
\address{School of Physics and Astronomy, University of St. Andrews, North Haugh, St. Andrews, Fife, KY16 9SS, Scotland}
\address{Department of Physics, Sofia University, James
Bourchier 5 Boulevard, 1164 Sofia, Bulgaria}
\author{Ferdinand Schmidt-Kaler}
\address{Institut f\"ur Quantenphysik, Johannes
Gutenberg-Universit\"at Mainz, 55099 Mainz, Germany}

\begin{abstract}
We study theoretically the collective E$\otimes$e Jahn-Teller-Dicke distortion in a system of trapped ions.
We focus in the limit of infinite range interactions in which an ensemble of effective spins interacts with two collective vibrational modes with U(1) symmetric couplings.
Our model is exactly solvable in the thermodynamical limit and it is amenable to be solved by exact numerical diagonalization for a moderate number of ions.
We show that trapped ions are ideally suited to study the emergence of spontaneous symmetry breaking of a continuous symmetry and magnetic structural phase transition in a mesoscopic system.
\end{abstract}

 \maketitle

\section{Introduction}\label{introduction}
%
Physical systems where bosonic modes interact with electronic or pseudospin degrees of freedom reveal a rich variety of phenomena in condensed matter and atomic physics.
A prominent example is given by the Jahn-Teller (JT) models \cite{Englman1972,Bersuker2006} which  describe the interaction of electronic orbital degrees of freedom with vibrational modes either in molecules or solids.
The JT effect is formulated as a structural instability of molecular configurations in electronically degenerate states.
In particular, the electron-phonon coupling shifts the potential minima of the nuclei, which leads to position reordering and molecular distortion. Similar to the molecular systems, the properties of some crystals are also strongly affected by the JT coupling, including symmetry breaking and structural phase transitions \cite{Englman1970}.
Furthermore, the strong electron-phonon coupling in the cooperative JT models is an important factor in the description of colossal magneto-resistance in manganites and high T$_{\rm c}$-superconductivity \cite{Millis1996,Tokura2000}.

Atomic systems such as ultracold atoms and trapped ions allow experimentalists
to implement JT models in a controllable way that is not possible in solid-state or molecular setups.
This is a motivation to push the current quantum technology toward the realization of Analogical Quantum Simulators (AQS).
The latter are controllable systems where interactions between particles can be tuned and quantum states can be accurately prepared and measured with high efficiency. Recently, physical realizations of JT couplings have been discussed in terms of two-level systems coupled to a bimodal cavity \cite{Larson2008} and Bose-Einstein condensates in the presence of spatially dependent laser fields \cite{Larson2009PRA}. These systems pave the way for studying a quantum phenomena such as ground-state entanglement \cite{Hines2004,Liberti2007} and creation of artificial non-Abelian magnetic fields \cite{Larson2009}. A quantum chaotic behavior in the energy spectrum of multi-spin lattice JT model was discussed in \cite{Majernikova2011}.

Among the most promising physical systems for implementing AQS are linear ion crystals interacting with external lasers or magnetic fields \cite{Schneider2012,Johanning2009}. The main advantages of trapped ions are their addressability, long coherence times and high fidelity readout.
The current available ion trapping technology allows to explore the physics of quantum phase transitions in complex spin systems
\cite{Porras2004,Friedenauer2008,Ivanov2011,Islam2011,Britton2012}, interacting bosons \cite{Porras2004BHM,Ivanov2009,Porras2012,Bermudez2012,Bermudez2010}, relativistic effects, \cite{Gerritsma2010,Bermudez2008} and quantum open systems \cite{Barreiro2011}.

\begin{figure}[tb]
\begin{center}
\includegraphics[width=0.6\linewidth]{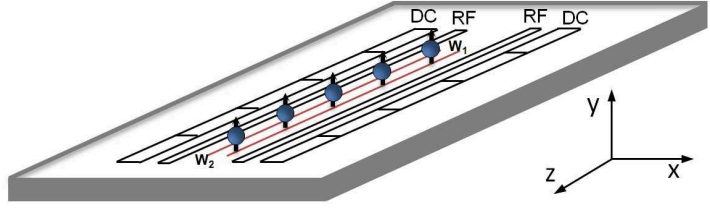}
\end{center}
\caption{Exemplary geometry for the realization of the collective Jahn-Teller-Dicke model with time-varying magnetic field gradient produced by
a surface-electrode trap. A linear ion crystal is oriented along the $z$ axis. Oscillating currents in two wires $W_{1}$ and $W_{2}$ parallel to the linear ion crystal are used to provide the symmetric spin-phonon coupling.} \label{fig1}
\end{figure}

In this work we propose an implementation of AQS of an
infinite range E$\otimes$e Jahn-Teller-Dicke (JTD) model based on a trapped ion crystal.
The doublet of electronic states is replaced here by two internal states of the ions.
The pair of molecular vibrational modes is represented by the two degenerate orthogonal center-of-mass (c.m) modes, respectively, in the two radial directions.
We show that the U(1) symmetric JTD spin-phonon coupling can be provided by applying a magnetic field with time-oscillating gradient, which couples the collective spin ensemble to the two orthogonal c.m modes \cite{Ospelkaus2008,JohanningPRL2009,Timoney2011}, Fig. \ref{fig1}. Such an oscillating magnetic field has been used to implement experimentally a single qubit rotation and multi-qubit quantum gates \cite{Ospelkaus2011}, which shows that our model could be realized with the current state-of-the art ion trap setup.
In a previous work \cite{Porras2012} we have focused on the cooperative JT model in which all vibrational modes are coupled to the effective spins, which leads to spin-phonon quasi-condensate. Here, we show that under suitable experimental conditions the spin ensemble may interact only with a single vibrational mode in each radial direction.
Thus, we deal with infinite range JT models, which are quasi-exactly solvable, in the sense that the ground state can be found in the thermodynamical limit \cite{EB} or even studied by exact numerical diagonalization with a moderate number of ions.

The E$\otimes$e JTD model possesses a continuous symmetry associated with rotation in the plane orthogonal to the trap axis.
There is a critical spin-phonon coupling above which the U(1) symmetry is spontaneously broken and the system evolves into one particular ground state which does not respect the same symmetry as the Hamiltonian. That broken symmetry is associated with a magnetic structural phase transition at zero temperature, where ions' equilibrium positions are \emph{arbitrarily} displaced in the radial $x$-$y$ plane together with the creation of macroscopic spin coherence. We show that the radial distortion of the ion crystal increases with the number of ions as $\sqrt{N}$, while the mean phonon number and the spin coherence scale as $N$, which is an analog to the normal-to-super-radiance phase transition in the Dicke model \cite{Baumann2010,Baumann2011}.

The paper is arranged as follows: In Sec. \ref{MDI} we describe the collective vibrational spectrum of the linear ion crystal. We show that an oscillating magnetic field gradient could select only one vibrational mode in each radial direction and thus to provide the symmetric JTD spin-phonon coupling. In Sec. \ref{MST} we explore the spin-phonon interaction assuming a thermodynamical limit. By using the well known technique of the Holstein-Primakoff representation we derive an analytical result for the amount of distortion and the spin ordering of the system. The experimental requirements for the physical implementation and readout of the final state of our model are discussed in Sec. \ref{EXP}.
Finally, in Sec. \ref{conclusion} we conclude our findings and present further interesting phenomena to be explored.
\section{The Trapped-Ion E$\otimes$e Jahn-Teller-Dicke model} \label{MDI}
%
\subsection{Trapped-ion radial vibrational Hamiltonian}
We consider a crystal of $N$ identical ions with mass $M$ and charge $q$ confined in a linear Paul trap along the $z$ axis.
Each ion has two metastable internal levels with energy separation $\tilde{\omega}_0$.
The system is described by the Hamiltonian ($\beta=x,y,z$ and $\hbar = 1$ from now on)
\begin{eqnarray}
H_{0} =H_{\rm spin}
+
H_{\rm vib},
\nonumber \\
H_{\rm spin}=\sum_{i=1}^{N}\frac{\tilde{\omega}_{0}}{2} \sigma_{i}^{z}, \quad H_{\rm vib} = \sum_{\beta} \sum_{i=1}^{N} \frac{p^2_{\beta,i}}{2 M}+V .
\end{eqnarray}
The first term in $H_0$ describes the energy of the two-level systems
with
$\sigma_{i}^{\beta}$ being the Pauli matrices for ion $i$.
$H_{\rm vib}$ is the vibrational Hamiltonian, which contains the ions' kinetic energy and the potential energy of the ion crystal.
The latter consists of the effective harmonic potential and the mutual Coulomb repulsion
\begin{equation}
V=\frac{M}{2} \sum_{\beta}\sum_{i=1}^{N}\omega^2_\beta r_{\beta,i}^2 +
\sum_{i>j}^{N} \frac{q^2}{|\vec{r}_i - \vec{r}_j|},\label{V}
\end{equation}
where $\vec{r}_{i}$ is the position vector operator of ion $i$ and $\omega_{\beta}$ denote the trapping frequencies.
In this work we consider the symmetry condition $\omega_x = \omega_y=\omega_{\rm r}$,
which can be achieved by proper adjustment of the trapping voltages or by special design of the trap geometry \cite{Singer2010}.
For sufficiently strong radial confinement ($\omega_{\rm r}\gg \omega_{z}$) ions occupied equilibrium positions $\vec{r}_{i0}=(0,0,z_{i}^{0})$ along the $z$ axis. The latter are determined by the balance between the Coulomb repulsion and the harmonic trapping force, which yields  $(\partial V/\partial \vec{r}_{i})_{\vec{r}_{i}=\vec{r}_{i0}}=0$. At low temperature the ions undergo only small oscillations around the equilibrium positions, namely
\begin{equation}
\vec{r}_{i}=\delta r_{x,i}\vec{e}_{x}+\delta r_{y,i}\vec{e}_{y}+(z_{i}^{0}+\delta r_{z,i})\vec{e}_{z},
\end{equation}
where $\delta r_{\beta,i}$ are the displacement operators.

The radial vibrational spectrum is essential for the implementation of our idea.
For that reason we discuss here its main characteristics, a more complete discussion can be found, for example in \cite{James1998,Deng2008}.
First we notice that a suitable length scale is given by $d_0^3 = q^2/(M \omega_z^2)$.
Accordingly, we define a dimensionless equilibrium positions, ${z'}_i^{0} = z_i^{0} / d_0$.
Making a Taylor expansion of the potential (\ref{V}) around ${z'}_i^{0}$ and neglecting $\delta r_{z,i}\delta r_{\alpha,i}^{2}$, $\delta r_{\alpha,i}^{3}$ and higher order terms, the radial vibration is described by the Hamiltonian ($\alpha=x,y$ from now on)
\begin{equation}
H_{\rm rad}=\sum_{\alpha}\sum_{i=1}^{N}\frac{p^2_{\alpha,i}}{2 M}+\frac{M\omega_{\rm r}^{2}}{2}\sum_{\alpha}\sum_{i,j=1}^{N}{\cal K}_{ij}
\delta r_{\alpha,i}\delta r_{\alpha,j}.
\end{equation}
Note that within harmonic approximation of the potential (\ref{V}) the radial motion is decoupled from the axial motion. The collective vibrational frequencies $\omega_{n}=\omega_{\rm r}\sqrt{\kappa_{n}}$ can found by solving the eigenvalue problem,
\begin{equation}
\sum_{j=1}^{N} {\cal K}_{ij} \ b_{j,n}^{\alpha} = \kappa_{n} \ b_{i,n}^{\alpha},\label{TrSpectrum}
\end{equation}
with the matrix ${\cal K}$ given by
\begin{equation}
{\cal K}_{ij} =
\left( 1 - \sum_{j'\neq j}^{N}
\frac{\omega_z^2}{\omega_{\rm r}^2} \frac{1}{|{z'}^0_j - {z'}^0_{j'}|^3} \right) \delta_{ij}
+ \frac{\omega_z^2}{\omega_{\rm r}^2} \frac{1}{|{z'}^0_i -{z'}^0_j|^3}
\left( 1 - \delta_{ij} \right).
\end{equation}
The equilibrium positions for the ions in natural units depend on the number of ions $N$ only, and thus, the radial vibrational modes are governed solely by the ratio $\omega_z / \omega_{\rm r}$.
\begin{figure}[tb]
\begin{center}
\includegraphics[width=0.6\linewidth]{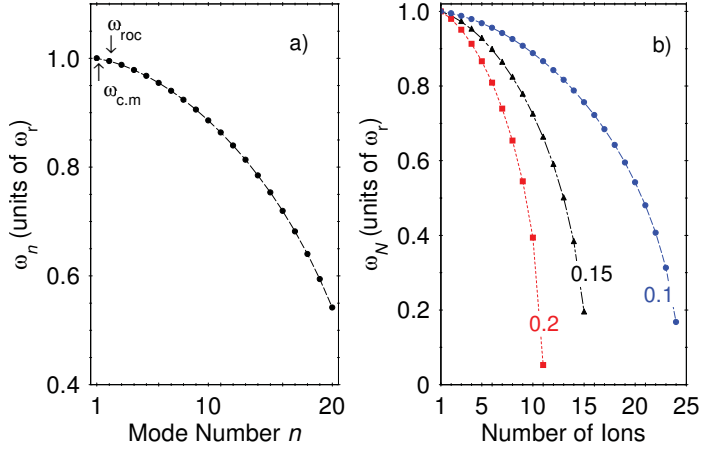}
\end{center}
\caption{Eigenfrequencies $\omega_{n}$ of the radial collective vibrational spectrum for a linear ion crystal with $N=20$ ions and $\omega_{z}/\omega_{\rm r}=0.1$.
The highest vibrational frequency is the $\rm c.m$ mode at $\omega_{\rm r}$.
The frequency splitting between $\omega_{\rm c.m}$ and the second-to-highest
rocking mode $\omega_{\rm roc}$ is independent of the number of ions.
b) The lowest energy vibrational frequency $\omega_{N}$ as a function of the number of ions for different
aspect ratio $\omega_{z}/\omega_{\rm r}=0.2,0.15$ and $0.1$.
As the number of ions $N$ is increased, a critical value occurs for which the
frequency $\omega_{N}$ becomes imaginary, consequently the ion crystal undergoes a structural transition into a zigzag phase. The critical $N$ varies
for different aspect ratio $\omega_{z}/\omega_{\rm r}$.} \label{fig2}
\end{figure}

The vibrational Hamiltonian $H_{\rm rad}$ can be diagonalized by defining
\begin{equation}
\delta r_{\alpha,i} = \sum_{n=1}^{N} b_{i,n}^{\alpha} q_{n}^{0}(a_{\alpha,n}^{\dag}+a_{\alpha,n}),
\ \
p_{\alpha,i} = {\rm i}\sum_{n=1}^{N} b_{i,n}^{\alpha} p_{n}^{0}(a_{\alpha,n}^{\dag}-a_{\alpha,n}) ,
\label{coll.coor}
\end{equation}
with $q_{n}^{0} = 1/\sqrt{2M\omega_{n}}$, and $p_{n}^{0} = \sqrt{M\omega_{n}/2}$. By substituting (\ref{coll.coor}) in $H_{\rm rad}$ we obtain a set of uncoupled collective vibrational modes, namely
\begin{equation}
H_{\rm rad}=\sum_{\alpha}\sum_{n=1}^{N}\omega_{n}\left(a_{\alpha,n}^{\dag}a_{\alpha,n}+\frac{1}{2}\right).
\end{equation}
One important feature of the radial vibrational modes is the fact that the highest two eigenvalues do not depend of $N$. In particular, if we order the vibrational eigenfrequencies by their energy, we find that the highest energy c.m mode
$\omega_{1} = \omega_{\rm c.m}=\omega_{\rm r}$ and second-to-highest energy rocking mode $\omega_{2} = \omega_{\rm roc} =
\sqrt{\omega_{\rm r}^2 - \omega_z^2}$ are independent of $N$, Fig. \ref{fig2}a. This is an important feature, which allows to resolve sidebands to the highest frequency $\omega_{\rm c.m}$ with a resolution independent of $N$.
Increasing $N$ the ion crystal undergoes a transition to a zigzag phase, which limits the number of ions one can achieve for a given aspect ratio $\omega_{z}/\omega_{\rm r}$ \cite{Fishman2008,Bermudez2012_2}. In Fig. \ref{fig2}b, we plot the evolution of the lowest energy radial mode $\omega_{N}$ with $N$, showing the maximum $N$ allowed for the linear configuration to be stable.
In the following we will show that the strong spin-phonon coupling in the JTD model induces a structural transition accompanied with magnetic ordering, which is a quantum version of the classical zigzag phase.

\subsection{Jahn-Teller {\rm E}$\otimes${\rm e} spin-phonon coupling}
We describe the interaction of the trapped ion spin ensemble with the collective vibrational modes.
Those interactions can be induced either by laser dipole forces or by magnetic field gradients. Here we focus on the latter technique since it naturally implements symmetric couplings in the $x$-$y$ plane, and it also avoids undesired effects like high-order terms in the Lamb-Dicke expansion and the spontaneous emission decoherence
\cite{JohanningPRL2009,Timoney2011}.
Let us assume that the ion crystal interacts with an oscillating magnetic quadrupole of the form
\begin{equation}
\vec{B}(t;x,y) = b f(t)(\vec{e}_{x}x-\vec{e}_{y}y).
\label{MF}
\end{equation}
Such a field can be created in a micro-structured planar ion trap, recently experimentally demonstrated \cite{Ospelkaus2011}, which contains two wires parallel to the linear ion crystal (Fig. \ref{fig1}) \cite{Welzel2011}. The magnetic field affects only the radial motion of the ion crystal and thus the motion along the $z$ axis can be safety neglected. We consider a time modulation $f(t)=(\cos\nu_{b}t+\cos\nu_{r}t)$
to control the couplings.  The magnetic dipole interaction is described by the interaction Hamiltonian
\begin{equation}
H_{\rm I} = - \sum_{i=1}^{N} \vec{\mu}_{i}\cdot\vec{B}(t;\delta r_{x,i}, \delta r_{y,i}),
\label{HI}
\end{equation}
where $\vec{\mu}_{i}=\mu_{x}\sigma_{i}^{x}+\mu_{y}\sigma_{i}^{y}$ is the magnetic dipole moment operator of the ion $i$, and we assume the condition $\mu_{x}=\mu_{y}=\mu$.
To control the spin-phonon couplings we choose driving frequencies
\begin{equation}
\nu_{b,r}=(\tilde{\omega}_0 - \omega_0)\pm(\omega_{\rm c.m}-\omega).
\label{FR}
\end{equation}
The goal is to drive spin-flip transitions with detuning $\omega_0$ as well as blue- and  red-sideband transitions of the c.m mode $\omega_{\rm c.m}$ with detuning $\pm \omega$.
The applied bichromatic magnetic field in $x$-$y$ plane, establishes Jaynes-Cummings and anti-Jaynes-Cummings interactions, which couple the internal and the motional states of the ions \cite{Ospelkaus2011,Molmer1999}.
The Hamiltonian in the interaction picture with respect to $H_{0}$ is given by
\begin{eqnarray}
H_{\rm I} &=&-\frac{\mu b}{2}\sum_{i=1}^{N}\delta r_{x,i}(t)(\sigma_{i}^{+}e^{{\rm i}\tilde{\omega}_{0} t}+\sigma_{i}^{-}e^{-{\rm i}\tilde{\omega}_{0}t})f(t)\nonumber \\
&&-{\rm i}\frac{\mu b}{2}\sum_{i=1}^{N}\delta r_{y,i}(t)(\sigma_{i}^{+}e^{{\rm i}\tilde{\omega}_{0}t} -\sigma_{i}^{-}e^{-{\rm i}\tilde{\omega}_{0} t})f(t),
\label{H1}
\end{eqnarray}
where $\sigma_{i}^{\pm}$ are the Pauli spin-flip operators. The displacement operators $\delta r_{\alpha,i}$ are recast in terms of collective operators by means of Eq. (\ref{coll.coor}) such that we can in a controlled way choose the driving frequencies to pick the radial c.m mode as the only resonant one. For this, the following set of conditions has to be satisfied,
\begin{eqnarray}
\omega_n \ll \tilde{\omega}_0, \nonumber \\
\lambda, \omega, \omega_0  \ll \omega_{\rm c.m} - \omega_{\rm roc},\label{condition}
\end{eqnarray}
where $\lambda=-\mu q_{0}b/\sqrt{2}$ is the spin-phonon coupling with $q_{0}\equiv q_{1}^{0}$ being the size of the c.m wave packet. The latter conditions ensures the approximation that any vibrational mode but the c.m one can be neglected in a rotating wave approximation. Consider as an example Zeeman $^{40}$Ca$^{+}$ qubits with transition frequency $\tilde{\omega}_{0}=20$ MHz confined in a planar trap with radial frequency $\omega_{\rm r}=2$ MHz, the first condition in Eq. (\ref{condition}) is justified. Assuming  $\omega_{z}/\omega_{\rm r}=0.2$ the frequency splitting is approximately $(\omega_{\rm c.m} - \omega_{\rm roc})\approx40$ kHz. With current ion-trap technology a spin-phonon coupling of order of $\lambda\approx5$ kHz is achieved by magnetic field gradient $b=25$ Tm$^{-1}$, which allowed to neglect the contribution of the off-resonant terms in Eq. (\ref{H1}). Under those assumptions we can approximate
the interaction Hamiltonian by%
\begin{eqnarray}
H_{\rm I} &=&
\frac{\lambda}{\sqrt{2N}}(a_{x}^{\dag}e^{{\rm i}\omega t}+a_{x}e^{-{\rm i}\omega t})(J_{+}e^{{\rm i}\omega_{0} t}+e^{-{\rm i}\omega_{0}t}J_{-})   \nonumber \\ &&+{\rm i}\frac{\lambda}{\sqrt{2N}}(a_{y}^{\dag}e^{{\rm i}\omega t}+a_{y}e^{-{\rm i}\omega t})(J_{+}e^{{\rm i}\omega_{0} t}-J_{-}e^{-{\rm i}\omega_{0}t}),
\label{H2}
\end{eqnarray}
Here $a_\alpha$ and $a_\alpha^\dagger$ correspond to the annihilation and creation operators of the c.m phonon, respectively.
Note that the factor $N^{-1/2}$ in (\ref{H2}) appears due to the excitation of the radial c.m modes, wherein the spin-phonon coupling scales as $b_{j,1}^{\alpha}=N^{-1/2}$.
Since the ions are equally coupled with the phonons we have introduced the collective spin operators $J_{+}=\sum_{i=1}^{N}\sigma_{i}^{+}$ ($J_{+}^{\dag}=J_{-}$) and $J_{z}=1/2\sum_{i=1}^{N}\sigma_{i}^{z}$, which describe the combined ionic pseudospin of length $j=N/2$. The collective spin basis is spanned by the Dicke states $|j,m\rangle $, which are eigenvectors of $J^{2}|j,m\rangle =j(j+1)|j,m\rangle$ and $J_{z}|j,m\rangle =m|j,m\rangle$, respectively. The Hilbert space of the total system is spanned by the states $\{|j,m\rangle\otimes|n_{x},n_{y}\rangle\}$, where $|n_{x,y}\rangle$ is the Fock state with $n_{x,y}$ phonons. After performing the time-dependent unitary transformation
$F
=
e^{{\rm i}\omega t(n_{x}+n_{y})+{\rm i}\omega_{0}tJ_{z} }$, such that $H_{\rm JTD}=F^{\dag}H_{\rm I}F-{\rm i}\hbar F^{\dag}\partial_{t}F$, we express the Hamiltonian (\ref{H2}) as
\begin{equation}
H_{\rm JTD}
=
\omega(n_{x}+n_{y})+\omega_{0}J_{z}
+
\frac{\lambda}{\sqrt{4j}}(J_{+}+J_{-})(a_{x}^{\dag}+a_{x})
+{\rm i}\frac{\lambda}{\sqrt{4j}}(J_{+}-J_{-})(a_{y}^{\dag}+a_{y}).
\label{H3}
\end{equation}
Hence we arrive at the realization of the collective JTD model, which describes a two-degenerate vibrational modes coupled to the effective spin ensemble by the symmetric JT coupling. The Hamiltonian (\ref{H3}) is a multi-particle extension of the E$\otimes$e model in molecular and solid-state physics. The trapped ion realization of JTD model allows for easy tuning of the effective spin and phonon frequencies by adjusting the detuning and the spin-phonon coupling via the magnetic gradient.

It is convenient to rewrite the Hamiltonian (\ref{H3}) in terms of right and left chiral operators \cite{Bermudez2008}
\begin{equation}
a_{r}^{\dag}=\frac{1}{\sqrt{2}}(a_{x}^{\dag}+{\rm i}a_{y}^{\dag}),\quad a_{l}=\frac{1}{\sqrt{2}}(a_{x}+{\rm i}a_{y}),\label{RL}
\end{equation}
which can be used to express the $z$ component of the total angular momentum $L_{z}=\sum_{j=1}^{N}L_{j}^{z}=n_{r}-n_{l}$.
Using (\ref{RL}), the Hamiltonian (\ref{H3}) is expressed in the form
\begin{eqnarray}
H_{\rm JTD}&=&\omega(a_{r}^{\dag}a_{r}+a_{l}^{\dag}a_{l})+\omega_{0}J_{z}
+\frac{\lambda}{\sqrt{2j}}J_{+}(a_{r}^{\dag}+a_{l}) +\frac{\lambda}{\sqrt{2j}}J_{-}(a_{r}+a_{l}^{\dag}),\label{HChiril}
\end{eqnarray}
which shows that in the JTD model the creation of collective atomic excitation is accompanied
by the creation (annihilation) of right (left) quantum of angular momentum and vice versa.
\begin{figure}[tb]
\begin{center}
\includegraphics[angle=0,width=0.5\linewidth]{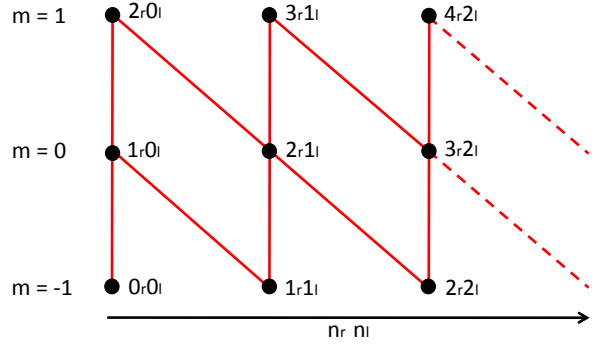}
\end{center}
\caption{Coupling pattern of the relevant ionic and vibrational states
$|j,m\rangle |n_{r},n_{l}\rangle$ for $j$=1. Due to the symmetries
in  spin-phonon coupling and parity of the Hamiltonian (\ref{H3}),
the quantum number $n_{r}-n_{l}-m$ is preserved and the Hilbert space is decomposed into subspaces of positive (negative) parity.
Here we show the non-vanishing couplings (solid lines) between states with $n_{r}-n_{l}-m=1$ and positive parity.} \label{fig3}
\end{figure}

\subsection{Symmetries}
Due to the symmetry in the spin-phonon interaction the JTD Hamiltonian (\ref{H3}) is invariant under the combined application
of a rotation in the $x$-$y$ plane
\begin{equation}
\left[
\begin{array}{ccc}
a_{x} \\
a_{y} \\
\end{array}%
\right]
=
\left[
\begin{array}{ccc}
\cos\phi & -\sin\phi \\
\sin\phi & \cos\phi \\
\end{array}%
\right]
\left[
\begin{array}{ccc}
a_{x}^{\prime} \\
a_{y}^{\prime} \\
\end{array}%
\right]
\end{equation}
and a phase shift $J_{+}\rightarrow e^{-{\rm i}\phi}J_{+}$. Hence, the JTD model is U(1) invariant, with the charge $C=L_{z}-J_{z}$
being the group generator, $[H,C]=0$. This implies that the Hilbert space is decomposed into subspaces with a well-defined quantum number $n_{r}-n_{l}-m$, Fig. (\ref{fig3}).
Because the Hamiltonian (\ref{H3}) is quadratic in the spin and phonon operators, it is also invariant under the application of the parity  operator $\Pi=\exp[{\rm i}\pi(n_{r}+n_{l}+J_{z}+j)]$. The Hilbert space of the total system is thus additionally decomposed into two noninteracting subspaces with even and odd number of total excitations \cite{Casanova2010}.

\subsection{Holstein-Primakoff representation}
In order to study the critical behavior of a collective JTD model (\ref{H3}) in the limit $j\rightarrow \infty$, we use the Holstein-Primakoff transformation, whereby the spin-$N/2$ degree of freedom is expressed in terms of single mode bosonic operators, namely $J_{+}=b^{\dag}\sqrt{2j-b^{\dag}b}$, $J_{-}=\sqrt{2j-b^{\dag}b}b$, and $J_{z}=b^{\dag}b-j$. This transformation preserves the spin algebra and allows to convert the JTD Hamiltonian (\ref{H3}) into the Hamiltonian
\begin{eqnarray}
H_{\rm JTD}&=&\omega(n_{x}+n_{y})+\omega_{0}(b^{\dag}b-j)+\frac{\lambda}{\sqrt{2}}\{b^{\dag}\sqrt{1-\frac{b^{\dag}b}{2j}}\nonumber \\
&&+\sqrt{1-\frac{b^{\dag}b}{2j}}b\}(a_{x}^{\dag}+a_{x})+{\rm i}\frac{\lambda}{\sqrt{2}}\{b^{\dag}\sqrt{1-\frac{b^{\dag}b}{2j}}\nonumber \\
&&-\sqrt{1-\frac{b^{\dag}b}{2j}}b\}(a_{y}^{\dag}+a_{y})\label{H_HP},
\end{eqnarray}
which describes three coupled bosonic field modes. This approach is the basis for the theoretical discussion in the following section.
\section{Magnetic Structural Phase Transition}\label{MST}
\begin{figure}[tb]
\begin{center}
\includegraphics[width=0.5\linewidth]{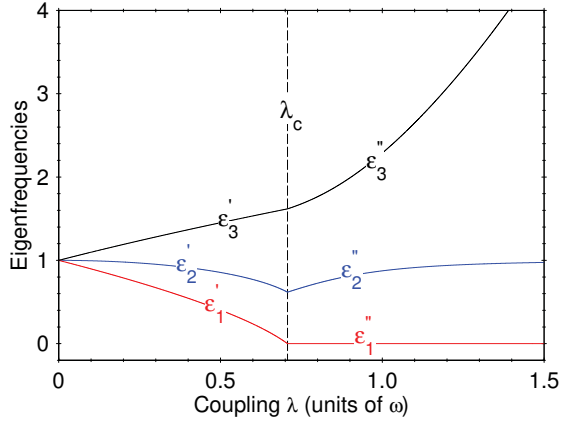}
\end{center}
\caption{The energy spectrum of the collective E$\otimes$e JTD model in the limit $j\rightarrow \infty$ as a function of the spin-phonon coupling $\lambda$ with $\omega=\omega_{0}$. For $\lambda\leq \lambda_{{\rm c}}$ the eigenfrequencies $\varepsilon_{p}^{\prime}$ are given as a solution of Eq. (\ref{EQ1}). At $\lambda=\lambda_{{\rm c}}$ the system undergoes a magnetic structural phase transition with broken U(1) symmetry. The new eigenfrequencies $\varepsilon_{p}^{\prime\prime}$ for $\lambda\geq \lambda_{{\rm c}}$ are given by Eq. (\ref{EQ2}).} \label{fig4}
\end{figure}

By taking the thermodynamical limit $j\rightarrow \infty$ the Hamiltonian (\ref{H_HP}) can be rewritten as follows
\begin{eqnarray}
H_{\rm JTD}&=&\omega(n_{x}+n_{y})+\omega_{0}(b^{\dag}b-j)+\frac{\lambda}{\sqrt{2}}(b^{\dag}+b)(a_{x}^{\dag}+a_{x})\nonumber \\
&&+{\rm i}\frac{\lambda}{\sqrt{2}}(b^{\dag}-b)(a_{y}^{\dag}+a_{y}),\label{HNorm}
\end{eqnarray}
which is bilinear in the bosonic field operators and therefore it can be exactly diagonalized by means of a Bogoliubov transformation, which yield (see \ref{appendix})
\begin{equation}
H_{\rm JTD}=\sum_{p=1}^{3}\varepsilon_{p}^{\prime}\left(c_{p}^{\dag}c_{p}+\frac{1}{2}\right)-\omega_{0}\left(j+\frac{1}{2}\right)-\omega\label{HD1}.
\end{equation}
The eigenfrequencies $\varepsilon_{p}^{\prime}$ can be found by solving the eigenvalue problem
\begin{equation}
\sum_{j=1}^{3}{\cal B}_{ij}^{\prime}v_{j}^{\prime(p)}=\varepsilon_{p}^{\prime 2}v_{i}^{\prime(p)}\label{EQ1}
\end{equation}
for the matrix
\begin{equation}
{\cal B}_{ij}^{\prime}=
\left[
\begin{array}{ccc}
\omega^{2} & -\lambda \sqrt{\frac{\omega_{0}\omega}{m_{+}}} & \lambda\sqrt{\frac{\omega_{0}\omega}{m_{-}}} \\
-\lambda \sqrt{\frac{\omega_{0}\omega}{m_{+}}} & \frac{\omega^{2}+\omega^{2}_{0}}{2m_{+}} & \frac{\omega^{2}-\omega^{2}_{0}}{2\sqrt{m_{+}m_{-}}} \\
\lambda\sqrt{\frac{\omega_{0}\omega}{m_{-}}} & \frac{\omega^{2}-\omega^{2}_{0}}{2\sqrt{m_{+}m_{-}}} & \frac{\omega^{2}+\omega^{2}_{0}}{2m_{-}}
\end{array}%
\right]\label{Bmatrix1},
\end{equation}
with $m_{\pm}=(1\pm\lambda\sqrt{2/\omega_{0}\omega})^{-1}$. The range of validity of the eigenfrequencies $\varepsilon_{p}^{\prime}$ is limited for $\lambda\leq\sqrt{\omega_{0}\omega/2}$. Indeed, the requirement for hermicity of the matrix (\ref{Bmatrix1}) is hold for $\lambda\leq\lambda_{{\rm c}}$, with $\lambda_{{\rm c}}=\sqrt{\omega_{0}\omega/2}$ being the critical coupling, Fig. \ref{fig4}. The vacuum state $\left\vert 0_{1}\right\rangle$ of Hamiltonian (\ref{HD1}) is defined by the condition $c_{p}\left\vert 0_{1}\right\rangle =0$. It is straightforward to show that the mean-value of the displacement of the c.m radial coordinates vanishes, $\langle \delta r_{x}\rangle=\langle \delta r_{y}\rangle=0$. The phase is characterized with zero phonon excitations and collective spin pointing along the $z$ axis,  $\langle J_{z}\rangle=-j$.

We may find a simple physical interpretation of the critical spin-phonon coupling $\lambda_{{\rm c}}$. Indeed, it is well known that in the presence of spin-orbit coupling, the minima of the lower adiabatic potential surface APS (effective nuclei potential in the molecular physics) for $\lambda\leq \lambda_{{\rm c}}$ appears at the origin, while for $\lambda>\lambda_{{\rm c}}$ APS has a sombrero shape. Increasing the spin-phonon coupling, the energy is minimized by breaking some spatial symmetry and thus leads to a JT distortion \cite{Bersuker2006}. In order to quantify amount of distortion and the spin ordering in the ion crystal above the critical coupling $\lambda_{{\rm c}}$, we follow the general procedure introduced by Emary and Brandes in \cite{EB} for the quantum Dicke model.
We displace each of the bosonic modes $a^{\dag}_{x}\rightarrow a^{\dag}_{x}+\sqrt{\alpha_{x}^{\ast}}$, $a^{\dag}_{y}\rightarrow a^{\dag}_{y}+\sqrt{\alpha_{y}^{\ast}}$, and $b^{\dag}\rightarrow b^{\dag}-\sqrt{\gamma^{\ast}}$, where $\alpha_{x}$, $\alpha_{y}$, and $\gamma$ are generally complex parameters in the order of $j$.
By using the Holstein-Primakoff representation and by substituting the displaced operators, the Hamiltonian (\ref{H_HP}) becomes
\begin{figure}[h]
\begin{center}
\includegraphics[width=0.6\linewidth]{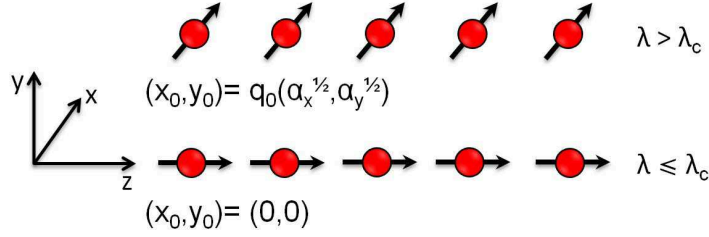}
\end{center}
\caption{A magnetic structural phase transition. For spin-phonon coupling $\lambda\leq\lambda_{{\rm c}}$ the ions' equilibrium positions are aligned along the $z$ axis, i.e. $x_{0}=y_{0}=0$ and the only non-zero projection of the total collective spin operator is $\langle J_{z}\rangle=-j$. Increasing $\lambda>\lambda_{{\rm c}}$ the system undergoes a magnetic structural phase transition, wherein the radial ions' equilibrium positions are displaced by $x_{0}=q_{0}\sqrt{\alpha_{x}}$ and $y_{0}=q_{0}\sqrt{\alpha_{y}}$, accompanied by a ferromagnetic spin ordering, $\langle J_{x,y}^{2}\rangle\neq 0$.  } \label{fig5}
\end{figure}
\begin{eqnarray}
H_{\rm JTD}&=&\omega(a^{\dag}_{x}a_{x}+\sqrt{\alpha_{x}^{\ast}}a_{x}+\sqrt{\alpha_{x}}a_{x}^{\dag}+\vert\alpha_{x}\vert)+\omega(a^{\dag}_{y}a_{y} \nonumber \\
&&+\sqrt{\alpha_{y}^{\ast}}a_{y}+\sqrt{\alpha_{y}}a_{y}^{\dag}+\vert\alpha_{y}\vert)+\omega_{0}(b^{\dag}b-\sqrt{\gamma^{\ast}}b  \nonumber \\
&&-\sqrt{\gamma}b^{\dag}+\vert\gamma\vert-j)+\frac{\lambda}{\sqrt{2}}\sqrt{\frac{k}{2j}}(a^{\dag}_{x}+a_{x}+\sqrt{\alpha_{x}^{\ast}} \nonumber \\
&&+\sqrt{\alpha_{x}})\{b^{\dag}\sqrt{\xi}+\sqrt{\xi}b-\sqrt{\xi}(\sqrt{\gamma^{\ast}}+\sqrt{\gamma})\}\nonumber \\
&&+{\rm i}\frac{\lambda}{\sqrt{2}}\sqrt{\frac{k}{2j}}(a^{\dag}_{y}+a_{y}+\sqrt{\alpha_{y}^{\ast}}+\sqrt{\alpha_{y}})\{b^{\dag}\sqrt{\xi} \nonumber \\
&&-\sqrt{\xi}b-\sqrt{\xi}(\sqrt{\gamma^{\ast}}-\sqrt{\gamma})\},
\label{HamD1}
\end{eqnarray}
where
\begin{equation}
k=2j-\vert\gamma\vert;\,\sqrt{\xi}=\sqrt{1-\frac{b^{\dag}b-b^{\dag}\sqrt{\gamma}-b\sqrt{\gamma^{\ast}}}{k}}.
\end{equation}
The parameters $\alpha_{x,y}$ and $\gamma$ can be found from the condition that all terms linear in the bosonic field operators in Eq. (\ref{HamD1}) are canceled
\begin{eqnarray}
\sqrt{\alpha_{x}}&=&\frac{\lambda}{\omega}\sqrt{j(1-s^{2})}\cos\phi,\,\sqrt{\alpha_{y}}=\frac{\lambda}{\omega}\sqrt{j(1-s^{2})}\sin\phi, \nonumber \\
\sqrt{\vert\gamma\vert}&=&\sqrt{j(1-s)},\label{parameters1}
\end{eqnarray}
with $s=\lambda_{{\rm c}}^{2}/\lambda^{2}$ and $\phi=\arg(\sqrt{\gamma})$, the second being arbitrary. The latter reflects the arbitrariness in the choice of a direction in spontaneous symmetry breaking, Fig. \ref{fig5}. Again, in the limit $j\rightarrow \infty$, the Hamiltonian (\ref{HamD1}) can be brought to the diagonal form (see \ref{appendix})
\begin{equation}
H_{\rm JTD}=\sum_{p=2}^{3}\varepsilon_{p}^{\prime\prime}\left(r_{p}^{\dag}r_{p}+\frac{1}{2}\right)-\omega-\frac{\omega_{0}}{4s}(1+s)-j\left(\frac{\lambda^{2}}{\omega}+\frac{\omega_{0}^{2}\omega}{4\lambda^{2}}\right)
-\frac{\lambda^{2}}{2\omega}(1-s).\label{HD2}
\end{equation}
The new frequencies $\varepsilon_{p}^{\prime\prime}$ are solution of the eigenvalue problem
\begin{equation}
\sum_{i=1}^{3}{\cal B}_{ij}^{\prime\prime}v_{i}^{\prime\prime(p)}=\varepsilon_{p}^{\prime\prime2}v_{j}^{\prime\prime(p)},\label{EQ2}
\end{equation}
for the matrix
\begin{equation}
{\cal B}_{ij}^{\prime\prime}=
\left[
\begin{array}{ccc}
\xi^{2}M_{-} & \lambda^{2}\sqrt{2M_{-}} & \nu\sqrt{M_{+}M_{-}} \\
\lambda^{2}\sqrt{2M_{-}} & \omega^{2} & -\lambda^{2}\sqrt{2M_{+}}  \\
\nu\sqrt{M_{+}M_{-}} & -\lambda^{2}\sqrt{2M_{+}}  & \xi^{2}M_{+}
\end{array}%
\right]\label{Bmatrix2},
\end{equation}
with $\xi^{2}=\left(\frac{\omega^{2}}{2}+\frac{\omega_{0}^{2}}{2s^{2}}\right)$, $\nu=\left(\xi^{2}-\frac{\omega_{0}^{2}}{s^{2}}\right)$ and $M_{\pm}=(1\pm s)$, respectively. In contrast to (\ref{EQ1}), now the frequencies $\varepsilon_{p}^{\prime\prime}$ remain positively defined in the region $\lambda\geq\lambda_{{\rm c}}$, Fig. \ref{fig4}.
\begin{figure}[tb]
\begin{center}
\includegraphics[width=0.5\linewidth]{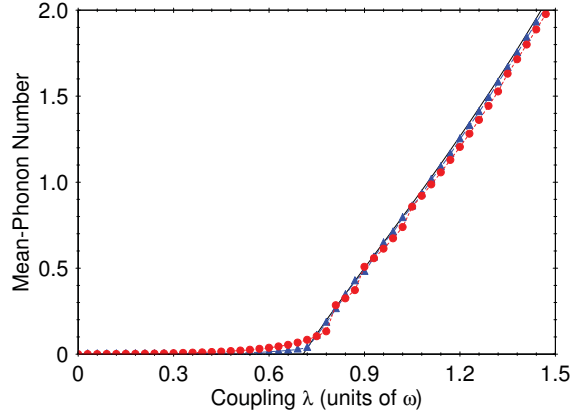}
\end{center}
\caption{The  mean-phonon number $(\langle n_{x}\rangle+\langle n_{y}\rangle)/j$ as a function
of the spin-phonon coupling $\lambda$. The numerical results for $10$ (red circles) and $20$ (blue triangles) ions are plotted together with the mean-field solution (black line).
For $\lambda \leq \lambda_{{\rm c}}$ the phase is characterized with zero mean-phonon number.
A position reordering is observed for $\lambda>\lambda_{{\rm c}}$ accompanied with non-zero mean-phonon number. The magnetic structural transition becomes sharper with increasing $N$.}\label{fig6}
\end{figure}

\begin{figure}[h]
\begin{center}
\includegraphics[width=0.5\linewidth]{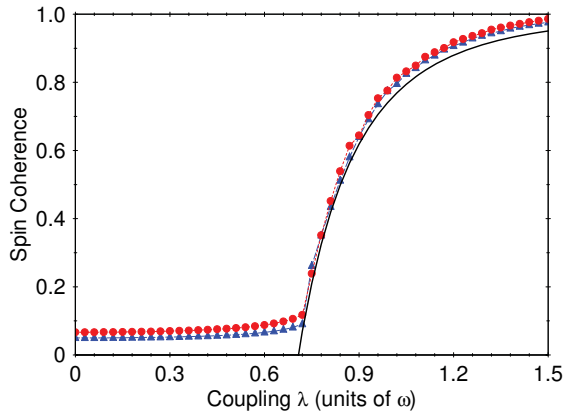}
\end{center}
\caption{The mean-field result for $(\langle J_{x}^{2}\rangle+\langle J_{y}^{2}\rangle)/j^{2}$ as a function of the spin-phonon coupling $\lambda$. The numerical results for 30 (red circles) and 40 (blue triangles) ions are compared with the mean-field solution. A creation of macroscopic spin-coherence is observed for $\lambda>\lambda_{{\rm c}}$, which is an analog to the super-radiance phase in the Dicke model.} \label{fig7}
\end{figure}

The mean-value phonon number with respect to the new vacuum state $\left\vert 0_{2}\right\rangle$ of the Hamiltonian (\ref{HD2}) with $r_{p}\left\vert 0_{2}\right\rangle=0$ is $\langle n_{x,y}\rangle=\alpha_{x,y}$, indicating a non-zero radial phonon excitations, Fig. \ref{fig6}. The collective displacement of the c.m mode implies a position reordering of the ions' equilibrium positions in the radial $x$-$y$ plane. Indeed, from Eq. (\ref{coll.coor}) it follows that the new radial equilibrium positions are $x_{0}=q_{0}\sqrt{\alpha_{x}}$ and $y_{0}=q_{0}\sqrt{\alpha_{y}}$. The structural transition also is accompanied with the ferromagnetic spin ordering, $\langle J_{x}^{2}\rangle=j^{2}(1-s^{2})\cos^{2}\phi$, $\langle J_{y}^{2}\rangle=j^{2}(1-s^{2})\sin^{2}\phi$, respectively, Fig. \ref{fig7}. We note that, the magnetic structural transition breaks the continuous U(1) symmetry of the JTD model, which reflects to the energy spectrum, namely one of the eigenfrequencies corresponds to the gapless Goldstone mode, $\varepsilon_{1}^{\prime\prime}=0$, see Fig. \ref{fig4}.
\section{Preparation and Detection of the Magnetic Structural Phase Transition}\label{EXP}
In the following we discuss the implementation of our model in a realistic trapped ion experiment. Consider an ion crystal which consists of $^{40}$Ca$^{+}$ ions with qubit states encoded at the Zeeman $S_{1/2}$ levels, $\left\vert \uparrow \right\rangle$ and $\left\vert \downarrow \right\rangle$. The experimental sequence is started by ground state laser cooling of the radial c.m modes and pumping spins to $\left\vert -j\right\rangle=\left\vert\downarrow \downarrow\ldots\downarrow\right\rangle$. Because the proposed method for implementation of our model is based on magnetic field interaction, the spectator modes could be only Doppler cooled. This is a key advantage compared to the laser-ion interaction, where the spin-phonon coupling would depend on the spectator modes by the Debye-Waller factor, which is a significant source of decoherence
\cite{Wunderlich2007}.
After the preparation of the initial state, the coupling $\lambda$ is slowly increased, relatively to the energy gap. For a moderate ion crystal with $N=10$ the energy gap around the critical coupling is approximately $\Delta E\approx0.25\omega$. The adiabatic condition requires $T\gg \Delta E^{-1}$, where $T$ is the interaction time. Assuming $\omega=\omega_{0}=4$ kHz, the adiabaticity is hold for $T\approx5$ ms which is comparable with the heating time \cite{Ospelkaus2011}. The magnetic structural phase transition can be detected by measuring either the radial displacement or the spin population at the end of the quantum simulation. Assuming  spin-phonon coupling $\lambda\approx 5$ kHz at the end of the quantum simulation and direction of spontaneous symmetry breaking $\phi=\pi/4$, the equilibrium positions are displaced in the radial $x$-$y$ directions by $x_{0}=y_{0}\approx 2q_{0}$. Such a structural transition can be detected by laser induced fluorescence, which is imaged on a CCD camera. The detection of the magnetic ordering can be performed by measuring the spin populations. For particular ion species, an illumination of the crystal with resonant light near $397$ nm and $866$ nm would provide a spin-dependent laser fluorescence, namely all spins up emit light and appear bright while spins down remain dark.

\section{Conclusion}\label{conclusion}
We have presented a proposal for the physical realization of the collective JTD model based on a linear ion crystal. We have shown that the JTD model exhibits a magnetic structural phase transition in the thermodynamical limit. Beyond the critical coupling the continuous U(1) symmetry is spontaneously broken which leads to collective motional displacement of the radial coordinates and creation of macroscopic spin-coherence. The features of the magnetic structural transition can be easily measured in the mesoscopic ion crystal by laser induced fluorescence. All parameters can be tuned by changing the detuning and the magnetic field gradient. In future we will investigate the JT effects in 2D ion crystals, which are relevant to orbital physics in solids.  Furthermore, the ion crystal also can serve as a platform for studying non-equilibrium phenomena and effects of decoherence in such many-body systems, which are computationally intractable.
\ack This work has been supported by the Bulgarian NSF grants D002-90/08, DMU-03/103, DMU-03/107, the EU 7th Framework Programme
collaborative project iQIT and RyC Contract No. Y200200074. P. A. Ivanov acknowledges the COST Action MP IOTA
1001.

\appendix

\section{Diagonalization of the Mean-Field Hamiltonian}\label{appendix}
\subsection{Normal Phase}
We start with the diagonalization of the Hamiltonian (\ref{HNorm}). It is convenient to work in the position-momentum representation by introducing position and momentum operators for each of the bosonic modes,
\begin{eqnarray}
&&a_{x}=\sqrt{\frac{\omega}{2}}x+\frac{{\rm i}}{\sqrt{2\omega}}p_{x},\quad a_{y}={\rm i}\sqrt{\frac{\omega}{2}}y-\frac{1}{\sqrt{2\omega}}p_{y},\nonumber \\
&&b=\sqrt{\frac{{\omega_{0}}}{2}}z+\frac{{\rm i}}{\sqrt{2\omega_{0}}}p_{z},\label{PM}
\end{eqnarray}
where the quantum oscillators have frequency $\omega$ in the $x$-$y$ plane and $\omega_{0}$ in the $z$ direction. The transformation gives
\begin{eqnarray}
H_{\rm JTD}&=&\frac{\omega^{2}}{2}(x^{2}+y^{2})+\frac{\omega_{0}^{2}}{2}z^{2}+\frac{1}{2}\left(p^{2}_{x}+p^{2}_{y}+p^{2}_{z}\right)  \nonumber \\
&&+\lambda\sqrt{2\omega_{0}\omega}z x-\lambda\sqrt{\frac{2}{\omega_{0}\omega}}p_{z}p_{y}-e_{0},\label{H4}
\end{eqnarray}
with $e_{0}=\omega_{0}(j+1/2)+\omega$. The effective Hamiltonian (\ref{H4}) describes a system of three quantum harmonic oscillators which are coupled through position and momentum dependent couplings. To express (\ref{H4}) as a set of uncoupled oscillators we need first to eliminate the momentum dependent interaction term. To achieve that first we rotate the coordinate system along the $x$ axis with the matrix
\begin{equation}
R_{x}=
\left[
\begin{array}{ccc}
1 & 0 & 0 \\
0 & \frac{1}{\sqrt{2}} & \frac{1}{\sqrt{2}} \\
0 & -\frac{1}{\sqrt{2}} & \frac{1}{\sqrt{2}}
\end{array}%
\right],
\end{equation}
such that $(x,y,z)^{T}=R_{x}(q_{1},q_{2},q_{3})^{T}$ and $(p_{x},p_{y},p_{z})^{T}=R_{x}(p_{1},p_{2},p_{3})^{T}$. The transformed Hamiltonian is given by
\begin{eqnarray}
H_{\rm JTD}&=&\frac{1}{2}(\omega^{2}q_{1}^{2}+p_{1}^{2})+\frac{1}{2}(\epsilon^{2}q_{2}^{2}+
\frac{p_{2}^{2}}{m_{+}})+\frac{1}{2}(\epsilon^{2}q_{3}^{2}+\frac{p_{3}^{2}}{m_{-}})\nonumber \\
&&+\lambda\sqrt{\omega_{0}\omega}q_{1}(q_{3}-q_{2})+(\omega^{2}-\epsilon^{2})q_{2}q_{3}-e_{0},\label{H5}
\end{eqnarray}
with $\epsilon^{2}=(\omega^{2}+\omega^{2}_{0})/2$. Hence, the momentum dependent coupling vanishes but as a consequence two of the effective harmonic oscillators acquire different effective masses
\begin{equation}
m_{+}=\left(1+\lambda\sqrt{\frac{2}{\omega_{0}\omega}}\right)^{-1},\quad m_{-}=\left(1-\lambda\sqrt{\frac{2}{\omega_{0}\omega}}\right)^{-1}.\label{M}
\end{equation}
The Hamiltonian (\ref{H5}) can be rewritten in a compact form as follows
\begin{equation}
H_{\rm JTD}=\frac{p_{1}^{2}}{2}+\frac{p_{2}^{2}}{2m_{+}}+\frac{p_{3}^{2}}{2m_{-}}+\frac{1}{2}\sum_{i,j=1}^{3}{\cal B}_{ij}^{ (1)}q_{i}q_{j}-e_{0}.\label{H6}
\end{equation}
Here ${\cal B}_{ij}^{(1)}$ is ($3\times 3$) real and symmetric matrix, given by
\begin{equation}
{\cal B}_{ij}^{(1)}=
\left[
\begin{array}{ccc}
\omega^{2} & -\lambda\sqrt{\omega_{0}\omega} & \lambda\sqrt{\omega_{0}\omega} \\
-\lambda\sqrt{\omega_{0}\omega} & \epsilon^{2} & \omega^{2}-\epsilon^{2} \\
\lambda\sqrt{\omega_{0}\omega} & \omega^{2}-\epsilon^{2} & \epsilon^{2}
\end{array}%
\right].
\end{equation}
Still, Hamiltonian (\ref{H6}) is not in the desirable normal mode form, because the quantum oscillators have different effective masses. To overcome this problem we normalize the position operators $q_{1}^{\prime}=q_{1}$, $q_{2}^{\prime}=\sqrt{m_{+}}q_{2}$, $q_{3}^{\prime}=\sqrt{m_{-}}q_{3}$ and momentum operators $p_{1}^{\prime}=p_{1}$, $p_{2}^{\prime}=p_{2}/\sqrt{m_{+}}$, $p_{3}^{\prime}=p_{3}/\sqrt{m_{-}}$, respectively, to obtain
\begin{equation}
H_{\rm JTD}=\frac{1}{2}\sum_{i=1}^{3}p_{i}^{\prime 2}+\frac{1}{2}\sum_{i,j=1}^{3}{\cal B}_{ij}^{\prime}q_{i}^{\prime}q_{j}^{\prime}-e_{0},
\end{equation}
where
\begin{equation}
{\cal B}_{ij}^{\prime}=
\left[
\begin{array}{ccc}
\omega^{2} & -\lambda \sqrt{\frac{\omega_{0}\omega}{m_{+}}} & \lambda\sqrt{\frac{\omega_{0}\omega}{m_{-}}} \\
-\lambda \sqrt{\frac{\omega_{0}\omega}{m_{+}}} & \frac{\omega^{2}+\omega^{2}_{0}}{2m_{+}} & \frac{\omega^{2}-\omega^{2}_{0}}{2\sqrt{m_{+}m_{-}}} \\
\lambda\sqrt{\frac{\omega_{0}\omega}{m_{-}}} & \frac{\omega^{2}-\omega^{2}_{0}}{2\sqrt{m_{+}m_{-}}} & \frac{\omega^{2}+\omega^{2}_{0}}{2m_{-}}
\end{array}%
\right].\label{B}
\end{equation}
To find the collective spin-phonon modes, we solve the eigenvalue problem
\begin{equation}
\sum_{j=1}^{3}{\cal B}_{ij}^{\prime}v_{j}^{\prime(p)}=\varepsilon_{p}^{\prime 2}v_{i}^{\prime(p)},
\end{equation}
for the eigenfrequencies $\varepsilon_{p}^{\prime}$ and the eigenvectors $\vec{v}^{\prime(p)}$ with $p=1,2,3$. Finally, we introduce a new set of bosonic field operators by the relation
\begin{equation}
q_{i}^{\prime}=\sum_{p=1}^{3}\frac{v_{i}^{\prime(p)}}{\sqrt{2\varepsilon_{p}^{\prime}}}(c_{p}^{\dag}+c_{p}),\quad p_{i}^{\prime}={\rm i}\sum_{p=1}^{3}\sqrt{\frac{\varepsilon_{p}^{\prime}}{2}}v_{i}^{\prime(p)}(c_{p}^{\dag}-c_{p}),
\end{equation}
and arrive at the diagonal Hamiltonian
\begin{equation}
H_{\rm JTD}=\sum_{p=1}^{3}\varepsilon_{p}^{\prime}\left(c_{p}^{\dag}c_{p}+\frac{1}{2}\right)-\omega_{0}\left(j+\frac{1}{2}\right)-\omega.\label{HDiag1}
\end{equation}

\subsection{Magnetic Structural Phase Transition}
In order to diagonalize Hamiltonian (\ref{HamD1}) in the limit $j\rightarrow \infty$ we perform the following two steps. $i)$ Expand the Hamiltonian (\ref{HamD1}) as a power series in $1/k$ and neglect the terms in order of $j$ in the denominator. $ii)$ Eliminate the terms in (\ref{HamD1}) which are linear in the bosonic operators by the condition, Eq. (\ref{parameters1}). The resulting Hamiltonian becomes
\begin{eqnarray}
H_{\rm JTD}&=&\omega(n_{x}+n_{y})+\frac{\omega_{0}}{2s}(1+s)b^{\dag}b+\frac{\lambda}{2}\sqrt{1+s}\times\nonumber \\
&&(a_{x}^{\dag}+a_{x})(b^{\dag}+b)+{\rm i}\frac{\lambda}{2}\sqrt{1+s}(a_{y}^{\dag}+a_{y})(b^{\dag}-b) \nonumber \\
&&-\frac{\lambda}{2}\frac{1-s}{\sqrt{1+s}}(e^{{\rm i}\phi}b^{\dag}+e^{-{\rm i}\phi}b)\{\cos\phi(a_{x}^{\dag}+a_{x}) \nonumber \\
&&+\sin\phi(a_{y}^{\dag}+a_{y})\}+\frac{\lambda^{2}}{4\omega}\frac{(1-s)(3+s)}{(1+s)}\times\nonumber \\
&&(e^{{\rm i}\phi}b^{\dag}+e^{-{\rm i}\phi}b)^{2}-\tilde{e}_{0},\label{HDispl1}
\end{eqnarray}
with
\begin{equation}
\tilde{e}_{0}=j\{\frac{\lambda^{2}}{\omega}+\frac{\omega_{0}^{2}\omega}{4\lambda^{2}}\}+\frac{\lambda^{2}}{2\omega}(1-s).
\end{equation}
We can further simplify (\ref{HDispl1}) by applying the following transformations
\begin{eqnarray}
a_{x}=d_{x}\cos\phi-d_{y}\sin\phi,\nonumber \\
a_{y}=d_{x}\sin\phi+d_{y}\cos\phi,
\end{eqnarray}
and $e^{{\rm i}\phi}b^{\dag}\rightarrow b^{\dag}$. Then, the Hamiltonian reads
\begin{eqnarray}
H_{\rm JTD}&=&\omega(d^{\dag}_{x}d_{x}+d^{\dag}_{y}d_{y})+\frac{\omega_{0}}{2s}(1+s)b^{\dag}b    \nonumber \\
&&+\frac{\lambda s}{\sqrt{1+s}}(b^{\dag}+b)(d^{\dag}_{x}+d_{x})+{\rm i}\frac{\lambda}{2}\sqrt{1+s}(b^{\dag}-b)(d^{\dag}_{y}+d_{y}) \nonumber
\\
&&+\frac{\lambda^{2}}{4\omega}\frac{(1-s)(3+s)}{(1+s)}(b^{\dag}+b)^{2}-\tilde{e}_{0}.\label{Hd}
\end{eqnarray}
Following the standard procedure, we introduce the position and momentum operators for each of the bosonic modes
\begin{eqnarray}
&&d_{x}=-{\rm i}\sqrt{\frac{\omega}{2}}X+\frac{1}{\sqrt{2\omega}}P_{X},\quad d_{y}=\sqrt{\frac{\omega}{2}}Y+\frac{{\rm i}}{\sqrt{2\omega}}P_{Y},\nonumber \\
&&b=-{\rm i}\sqrt{\frac{\tilde{\omega}}{2}}Z+\frac{1}{\sqrt{2\tilde{\omega}}}P_{Z},
\end{eqnarray}
with $\tilde{\omega}=(\omega_{0}/2s)(1+s)$. The Hamiltonian (\ref{Hd}) in the position-momentum representation is given by
\begin{eqnarray}
H_{\rm JTD}&=&\frac{\omega^{2}}{2}(X^{2}+Y^{2})+\frac{\tilde{\omega}^{2}}{2}Z^{2}+\frac{1}{2}(P_{X}^{2}+P_{Y}^{2})+\frac{P_{Z}^{2}}{2m}
\nonumber \\
&&+\frac{2s}{1+s}P_{Z}P_{X}-\lambda^{2}(1+s)Z Y\nonumber \\
&&-\omega-\frac{\tilde{\omega}}{2}-\tilde{e}_{0},
\end{eqnarray}
with $m=(1+s)^{2}/4$. It is convenient to normalize the position and momentum operators in $x$ and $y$ directions as follows
\begin{eqnarray}
&&\sqrt{m}P_{X}\rightarrow P_{X},\quad \frac{X}{\sqrt{m}}\rightarrow X,\nonumber \\
&&\sqrt{m} P_{Y}\rightarrow P_{Y},\quad \frac{Y}{\sqrt{m}}\rightarrow Y.
\end{eqnarray}
Then the Hamiltonian reads
\begin{eqnarray}
H_{\rm JTD}&=&\frac{m\omega^{2}}{2}(X^{2}+Y^{2})+\frac{\tilde{\omega}^{2}}{2}Z^{2}\nonumber \\
&&+\frac{1}{2m}(P_{X}^{2}+P_{Y}^{2}+P_{Z}^{2})+\frac{s}{m}P_{Z}P_{X}-2\lambda^{2}m Z Y\nonumber \\
&&-\omega-\frac{\tilde{\omega}}{2}-\tilde{e}_{0}.\label{H123}
\end{eqnarray}
Similar as before, the diagonalization proceeds by nullifying the momentum dependent interaction term in (\ref{H123}), which is achieved by rotating the coordinate system along the $y$ axis with the matrix
\begin{equation}
R_{y}=
\left[
\begin{array}{ccc}
\frac{1}{\sqrt{2}} & 0 & \frac{1}{\sqrt{2}} \\
0 & 1 & 0 \\
-\frac{1}{\sqrt{2}} & 0 & \frac{1}{\sqrt{2}}
\end{array}%
\right],
\end{equation}
such that $(X,Y,Z)^{T}=R_{y}(Q_{1},Q_{2},Q_{3})^{T}$ and $(P_{X},P_{Y},P_{Z})^{T}=R_{y}(P_{1},P_{2},P_{3})^{T}$.
After performing the rotation we obtain
\begin{eqnarray}
H_{\rm JTD}&=&\frac{m\xi^{2}}{2}Q_{1}^{2}+\frac{m\omega^{2}}{2}Q_{2}^{2}+\frac{m\xi^{2}}{2}Q_{3}^{2}\nonumber \\
&&+\frac{P_{1}^{2}}{2m_{1}}+\frac{P_{2}^{2}}{2m}+\frac{P_{3}^{2}}{2m_{3}}+m\nu Q_{1}Q_{3}\nonumber \\
&&-\sqrt{2}m\lambda^{2}Q_{2}(Q_{3}-Q_{1})-\omega-\frac{\tilde{\omega}}{2}-\tilde{e}_{0},
\end{eqnarray}
with $\xi^{2}=\left(\frac{\omega^{2}}{2}+\frac{\omega_{0}^{2}}{2s^{2}}\right)$ and $\nu=\left(\xi^{2}-\frac{\omega_{0}^{2}}{s^{2}}\right)$, respectively. Two of the quantum oscillators acquire new effective masses
\begin{equation}
m_{1}=\frac{m}{M_{-}},\quad m_{3}=\frac{m}{M_{+}},
\end{equation}
with $M_{\pm}=(1\pm s)$. Again as before to express the Hamiltonian (\ref{H6}) in normal mode form we normalize the position and momentum operators, namely: $Q_{1}^{\prime \prime}=\sqrt{m_{1}}Q_{1}$, $Q_{2}^{\prime \prime}=\sqrt{m}Q_{2}$ and $Q_{3}^{\prime \prime}=\sqrt{m_{3}}Q_{3}$ and, respectively, $P_{1}^{\prime \prime}=P_{1}/\sqrt{m_{1}}$, $P_{2}^{\prime \prime}=P_{2}/\sqrt{m}$ and $P_{3}^{\prime \prime}=P_{3}/\sqrt{m_{3}}$. Then the Hamiltonian becomes
\begin{equation}
H_{\rm JTD}=\frac{1}{2}\sum_{i=1}^{3}P_{i}^{\prime \prime 2}+\frac{1}{2}\sum_{i,j=1}^{3}{\cal B}_{ij}^{\prime \prime}Q_{i}^{\prime \prime}Q_{j}^{\prime \prime}-\omega-\frac{\tilde{\omega}}{2}-\tilde{e}_{0},\label{HNorm2}
\end{equation}
where
\begin{equation}
{\cal B}_{ij}^{\prime \prime}=
\left[
\begin{array}{ccc}
\xi^{2}M_{-} & \lambda^{2}\sqrt{2M_{-}} & \nu\sqrt{M_{+}M_{-}} \\
\lambda^{2}\sqrt{2M_{-}} & \omega^{2} & -\lambda^{2}\sqrt{2M_{+}}  \\
\nu\sqrt{M_{+}M_{-}} & -\lambda^{2}\sqrt{2M_{+}}  & \xi^{2}M_{+}
\end{array}%
\right]\label{B2}.
\end{equation}
The new eigenfrequencies are obtained by solving the eigenvalue problem,
\begin{equation}
\sum_{j=1}^{3}{\cal B}_{ij}^{\prime \prime}v_{j}^{\prime\prime(p)}=\varepsilon_{p}^{\prime \prime 2}v_{i}^{\prime\prime(p)}.
\end{equation}
We find that one eigenvalue $\varepsilon_{1}^{\prime \prime}=0$ corresponds to a free mode. The latter is the Goldstone mode related to the breaking of the U(1) symmetry. We may define bosonic creation $r_{p}^{\dag}$ and annihilation $r_{p}$ operators for the nonzero energy modes $\varepsilon_{2,3}^{\prime \prime}$ by the relation
\begin{equation}
Q_{i}^{\prime \prime}=\sum_{p=2}^{3}\frac{v_{i}^{\prime\prime(p)}}{\sqrt{2\varepsilon_{p}^{\prime \prime}}}(r_{p}^{\dag}+r_{p}),\quad P_{i}^{\prime \prime}={\rm i}\sum_{p=2}^{3}\sqrt{\frac{\varepsilon_{p}^{\prime \prime}}{2}}v_{i}^{\prime\prime(p)}(r_{p}^{\dag}-r_{p}),\label{QP}
\end{equation}
Submitting (\ref{QP}) in (\ref{HNorm2}) we obtain the following diagonal Hamiltonian, which refers to two decoupled oscillators,
\begin{equation}
H_{\rm JTD}=\sum_{p=2}^{3}\varepsilon_{p}^{\prime\prime}\left(r_{p}^{\dag}r_{p}+\frac{1}{2}\right)-\omega-\frac{\omega_{0}}{4s}(1+s)-j\left(\frac{\lambda^{2}}{\omega}+
\frac{\omega_{0}^{2}\omega}{4\lambda^{2}}\right)-\frac{\lambda^{2}}{2\omega}(1-s).
\end{equation}

\end{document}